\documentclass[aps,prl,twocolumn,floats,floatfix,superscriptaddress,nofootinbib]{revtex4}
\usepackage{graphicx,color}
\usepackage{amssymb,amsmath}
\usepackage{tabularx}
\def\vec#1{\mbox{\boldmath $#1$}}

\begin{document}

\title{Imaginary-time theory for triple-alpha reaction rate}

\author{T. Akahori}
\affiliation{
Institute of Physics, University of Tsukuba,
Tsukuba 305-8571, Japan}
\author{Y. Funaki}
\affiliation{
Nishina Center for Accelerator-Based Science, 
The Institute of Physical and Chemical Research (RIKEN), 
Wako 351-0198, Japan}
\author{K. Yabana}
\affiliation{
Center for Computational Sciences,
University of Tsukuba, Tsukuba 305-8571, Japan }
\affiliation{
Institute of Physics, University of Tsukuba,
Tsukuba 305-8571, Japan}

\begin{abstract}
Using imaginary-time theory, it is shown that the triple-alpha reaction rate can be reliably calculated without the need to solve 
scattering problems involving three charged particles.
The calculated reaction rate is found to agree well with the 
empirical NACRE rate, which is widely adopted in stellar 
evolution calculations. The reason for this is explained using $R$-matrix theory. Extremely slow convergence is found to occur when a coupled-channel expansion is introduced, 
which helps to explain the very different reaction rates obtained using different theoretical approaches.
\end{abstract}

\maketitle

The triple-alpha reaction is a key process that influences the production of all heavy 
elements in the universe. 
Accurate knowledge of the reaction rate is essential for 
understanding stellar evolution and nucleosynthesis. 
Since experimental measurements are not feasible for this 
reaction, theoretical evaluation of the reaction rate is 
crucially important. 

In the triple-alpha process, the importance of 
$^{12}$C and $^8$Be resonances is well recognized \cite{Sa52,Ho54}.
At high temperature, the reaction proceeds dominantly 
through a resonant $0^+$ state of $^{12}$C at 7.65 MeV, 
which is known as the Hoyle state. At lower temperatures, 
processes that do not involve the Hoyle state become 
important. An empirical reaction rate assuming successive two-body
reactions of $\alpha$-$\alpha$ and $\alpha$-$^8$Be has
been derived \cite{No85,La86,De87}, and is adopted in the
NACRE compilation \cite{An99} as the standard rate to be used
in stellar evolution calculations.

However, the validity of the empirical rate formula should be 
confirmed by calculations based on microscopic quantum 
theories. Several theoretical attempts to calculate the rate 
using quantum theory involving three $\alpha$-particles have 
recently been undertaken. The first was conducted by 
Ogata and coworkers \cite{Og09}, and employed 
continuum-discretized coupled-channel (CDCC) theory, 
which is a well-established theory for direct nuclear reactions \cite{Ka86}. 
A surprisingly high value for the reaction rate was found
at low temperatures, and at $T=0.01$ GK it was larger than the NACRE value by 26 orders of magnitude \cite{An99}. 
Soon after this report was published, the consequences of the new rate
for the present understanding of stellar evolution were investigated
\cite{Do09,Sa10,Su11}. It was pointed out that such a high 
rate would not be compatible with the standard picture of 
stellar evolution; for example, the red giant phase 
disappears if the rate is adopted \cite{Do09,Su11}.
Following the report by Ogata et al., calculations using 
different quantum three-body approaches 
have been carried out \cite{Ga11,Is12,Ng12,Is13,Ng13,Su13}.
Unfortunately, there is a large degree of scatter in the reported rates at low temperatures, which vary between the NACRE rate \cite{An99} and that determined by the CDCC calculation \cite{Og09}. 
In view of the number of successful achievements of nuclear three-body 
reaction theories, this huge discrepancy among the reported 
rates is both surprising and puzzling.
However, two possible explanations can be put forward. The first is the lack of a rigorous scattering theory for three charged particles. The second is 
related to the quantum-tunneling nature of the process: 
the $\alpha$-particle travels through the Coulomb barrier  over a 
long distance, typically a few hundred femtometers, 
causing the reaction rate to be extremely small, and to vary by 60 orders of magnitude  within the range of astrophysically relevant temperatures.

Recently, we have proposed a new theoretical approach for 
determining the radiative capture reaction rate, which we refer to as 
imaginary-time theory \cite{Ya12}. In this theory, imaginary time is identified with inverse temperature as
is often used in quantum many-body theories of nonequilibrium systems. A related approach has been developed for the theory of
chemical reaction rates \cite{Se92}. 
In this letter, we report the application of imaginary-time
theory to the determination of the triple-alpha reaction rate. Since the theory 
does not require any scattering problems to be solved, 
it is ideally suited to the triple-alpha process, for which 
there is no formal scattering theory available. Indeed, it will be shown that the reaction rate can
be reliably calculated without any numerical problems.
The calculated rate is found to be virtually identical 
to the empirical NACRE rate, and no enhancement occurs at low temperatures. The reason for this good agreement is investigated analytically by combining
$R$-matrix theory~\cite{La58} with imaginary-time theory.

The following expression describes the triple-alpha 
reaction rate \cite{Ya12},
\begin{eqnarray}
&&N_{\rm A}^2\langle \alpha\alpha\alpha \rangle
=
6 \cdot 3^{3/2} N_{\rm A}^2 \left( \frac{2\pi\beta\hbar^2}{M_{\alpha}} \right)^3
\frac{8\pi(\lambda+1)}{\hbar\lambda((2\lambda+1)!!)^2}
\nonumber\\
&&
\times \sum_{M_f \mu} \langle \Phi_f \vert M_{\lambda\mu}
e^{-\beta H} \left( \frac{H-E_f}{\hbar c} \right)^{2\lambda+1}
P M_{\lambda\mu}^{\dagger} \vert \Phi_f \rangle
\label{rate-3a}
\end{eqnarray}
where $\beta = 1/k_B T$ is the inverse temperature,
$M_{\alpha}$ is the mass of an $\alpha$-particle,
$H$ is the Hamiltonian for three $\alpha$ particles, 
$\Phi_f$ is the wave function for $^{12}$C in the final state 
after $\gamma$-ray emission, $E_f$ is the energy of the 
final state measured from the three $\alpha$ threshold, 
and $M_f$ is the magnetic quantum number for the final state.
$M_{\lambda\mu}$ is the multipole transition operator for 
$\gamma$-ray emission with a multipolarity $\lambda$. 
The reaction of three $\alpha$-particles with total angular momentum $J=0$ is considered, which leads to the emission of a
$\gamma$-ray with $\lambda=2$, and the $2^+$ state of $^{12}$C at 4.44 MeV
for the final wave function $\Phi_f$.
$P$ is the projection operator which eliminates any bound
eigenstates of the three-body Hamiltonian.
It can easily be shown that Eq.~(\ref{rate-3a}) exactly coincides 
with the expression for the triple-alpha reaction rate, 
Eqs.~(2) $\sim$ (5) of \cite{de11}, by inserting a completeness 
relation for a functional space of three $\alpha$-particles into
Eq.~(\ref{rate-3a}).

In practice, to evaluate the reaction rate using Eq.~(\ref{rate-3a}), 
it is necessary to calculate
$\Psi(\beta)=e^{-\beta H} M^{\dagger}_{\lambda\mu}\Phi_f$.
This is achieved by evolving the wave function along the 
imaginary-time axis,
\begin{equation}
-\frac{\partial}{\partial \beta} \Psi(\beta) = H \Psi(\beta),
\label{iteq}
\end{equation}
starting with the initial wave function,
$\Psi(\beta=0)=M^{\dagger}_{\lambda\mu}\Phi_f$.
The reaction rate at an inverse temperature $\beta$ is then 
evaluated using the wave function at $\beta/2$,
\begin{equation}
\langle \alpha\alpha\alpha \rangle \propto
\sum_{M_f \mu} 
\left\langle \Psi \left(\frac{\beta}{2} \right)\right\vert 
\left(\frac{H-E_f}{\hbar c}\right)^5 
\left\vert\Psi \left(\frac{\beta}{2} \right) \right\rangle.
\end{equation}

The numerical calculations are carried out using the model
space and the three-body Hamiltonian described below.
In the calculations, the $\alpha$-particles are treated as point particles. The assumption of
dominant $J=0$ contribution is expected to be valid below $T < 1.0$ GK~\cite{de10,de11}. 
A Jacobi coordinate system is used, defined by 
$\vec r = \vec r_1 - \vec r_2$ and 
$\vec R = (\vec r_1 + \vec r_2)/2 - \vec r_3$, where
$\vec r_i$ $(i=1-3)$ are the coordinates of the three $\alpha$-particles.
The three-body wave function is expanded in partial waves,
$\Psi(\beta) = \sum_L (u_L(R,r,\beta)/Rr)
[Y_L(\hat R) Y_L(\hat r)]_{J=0}$. In the present work, 
only the $L=0$ component is considered, since this is expected to be the most 
important at low temperature. Ogata et al. adopted the same model 
space for their CDCC calculations \cite{Og09}. 
The radial variables $R$ and $r$ are discretized with a 
grid size $\Delta R=\Delta r=0.5$ fm, and radial 
grid points are employed up to the maximum values, $R_{\rm max}$ and $r_{\rm max}$.
The differential operators in the Hamiltonian are treated
using a nine-point finite difference formula. To solve 
Eq.~(\ref{iteq}), the Taylor expansion method is used for
short-time evolution with a step size of 
$\Delta \beta = 0.004$ MeV$^{-1}$.

The Hamiltonian $H$ for the three $\alpha$-particles is
constructed as follows. For the potential between two 
$\alpha$-particles, the Ali-Bodmer potential is used, considering only 
the $l=0$ angular momentum channel \cite{Al66}. 
The potential parameter is modified slightly so that 
it accurately reproduces the resonance corresponding to the 
ground state of $^8$Be at 92.08 keV. 
A three-body potential among the three $\alpha$-particles is added, and is given by
$V_{3\alpha}(\vec r_1,\vec r_2,\vec r_3)=
V_3 e^{-\mu(r_{12}^2+r_{23}^2+r_{31}^2)}$
with $\mu=0.15$ fm$^{-2}$. The value of $V_3$ is chosen
so that the resonance energy of the Hoyle state, 
the $0_2^+$ state of $^{12}$C, is reproduced at 379.8 keV 
above the three $\alpha$ threshold.
The final wave function $\Phi_f$ for $^{12}$C 
$J^{\pi}=2^+$ at 4.44 MeV is constructed 
using the orthogonality condition model~\cite{OCM}.

\begin{figure}[t]
 \centering
 \includegraphics[scale=0.7]{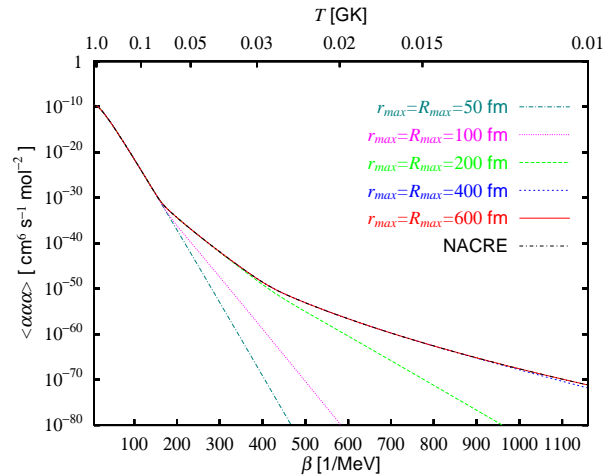}
 \caption{(Color online)
Calculated triple-alpha reaction rates for different 
choices of radial cutoff distances, $r_{\rm max}$ and $R_{\rm max}$.
The NACRE rate is also shown for comparison.}\label{fig:rateNr}
\end{figure}

In Fig.~\ref{fig:rateNr}, the calculated triple-alpha reaction rates 
for different spatial areas specified by $R_{\rm max}$ and 
$r_{\rm max}$ are compared. The NACRE rate \cite{An99} 
is also shown for comparison. It can be seen that when $R_{\rm max}$ and $r_{\rm max}$ are larger than 400 fm, a fully converged reaction rate
is obtained in the entire temperature region, and this rate coincides well with the NACRE rate.
Calculations within smaller spatial areas yield a rate which is 
valid only in limited higher-temperature regions. 
It has been previously shown that for a two-body radiative
capture process, the maximum radius required for full convergence 
roughly corresponds to the exit point 
of quantum tunneling at the Gamow peak energy \cite{Ya12}.
Using $R_{\rm max}=r_{\rm max}=400$ fm at $T=0.01$ GK,
the Gamow peak energy for the triple-alpha 
reaction can be estimated from the Coulomb potential, and is given by
$E_0 \simeq 3 \times 4e^2/400{\rm fm} = 43$ keV.

Although the calculated reaction rate and the NACRE
rate show good agreement on the logarithmic scale in
Fig. \ref{fig:rateNr}, the calculated rate is smaller than the NACRE
rate over the entire temperature region by a factor of up to six times. The reason for this difference will be discussed later.

In the empirical NACRE formula, there are three temperature regions 
that are distinguished by different reaction mechanisms \cite{No82}: 
$T > 0.074$ GK dominated by the Hoyle state process, 
0.074 GK $> T >$ 0.028 GK dominated by the $\alpha$-$^8$Be 
two-body nonresonant process, and $T < 0.028$ GK dominated 
by a nonresonant process involving three $\alpha$-particles. 
A careful look at Fig.~\ref{fig:rateNr} reveals that the calculated rate 
curves show changes in slope at exactly the same 
temperatures. 

\begin{figure}[tb]
 \centering
 \includegraphics[scale=0.7]{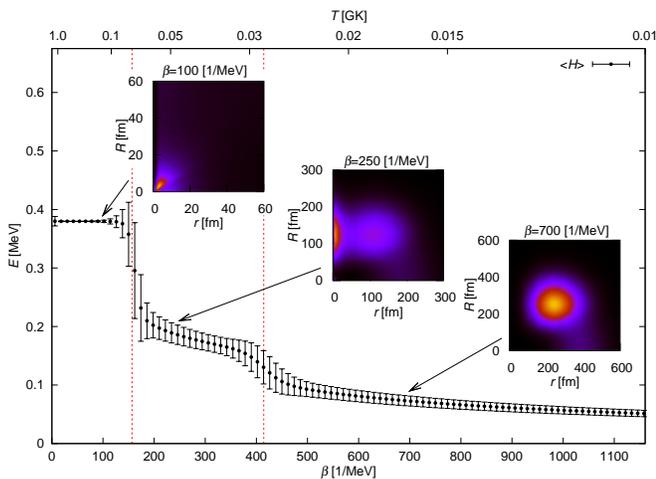}
 \caption{(Color online) 
Energy expectation value and variance as function
of temperature. The insets show density distributions for three different temperatures.}
\label{fig:GamovE}
\end{figure}

To illustrate this more clearly, Figure \ref{fig:GamovE} shows
the energy expectation value and the variance, defined by
${\bar E}= 
\langle \Psi(\beta/2) \vert H \vert \Psi(\beta/2) \rangle
/\langle \Psi(\beta/2) \vert \Psi(\beta/2) \rangle$
and 
$(\Delta E)^2 =
\langle \Psi(\beta/2) \vert (H-{\bar E})^2 \vert
\Psi(\beta/2) \rangle
/\langle \Psi(\beta/2) \vert \Psi(\beta/2) \rangle$,
as a function of the inverse temperature, respectively. 
The insets show the density distribution, 
$\rho(R,r,\beta/2)=u_{L=0}(R,r,\beta/2)^2/R^2r^2$,
for three typical temperatures.

In the high temperature region $T > 0.074$ GK, the energy expectation 
value coincides with the resonance energy of the Hoyle state, 
${\bar E}=379.8$ keV, indicating the dominance of the Hoyle 
state process. In addition, the density $\rho(R,r,\beta/2)$ is localized within a 
small $R$ and $r$ region, which is consistent with the resonant picture.
In the medium temperature region, 0.074 GK $> T >$ 0.028 GK, 
most of the density is contained within a small $r$ region, $r<10$ fm, 
whereas it is extended along the $R$ direction. This indicates that two of the $\alpha$-particles
are forming a $^8$Be resonance, with the third remaining 
outside. In the lowest temperature region, $T < 0.028$ GK, 
the density extends in both the $R$ and $r$ directions, indicating 
the nonresonant character of the reaction.
Note that the average energy at $T=0.01$ GK is $\bar E=55$ keV, which is in reasonable agreement with the Gamow peak energy of $ 43$ keV estimated from the radial
convergence shown in Fig.~\ref{fig:rateNr}.

The agreement between the calculated and NACRE rates, 
not only in terms of the magnitude, but also with regard to the change in reaction 
mechanism, indicates that imaginary-time theory provides 
quantum-mechanical support for the conventional description. 
An analytic investigation based on microscopic 
three-body theory was next carried out to determine if the empirical 
formula \cite{An99,No85} is justified. 

In the empirical formula, it is assumed that successive $\alpha$-$\alpha$ and $\alpha$-$^8$Be reactions occur, and the reaction cross sections are
described using Breit-Wigner formulas.
It will be shown that it is possible to derive a formula quite similar
to the empirical one starting from Eq.~(\ref{rate-3a}), by
assuming that the three-body Hamiltonian is separable, and then approximating it using 
$R$-matrix theory \cite{La58}.

The separability assumption for the three-body Hamiltonian
is written as
\begin{equation}
H=h_{\alpha\alpha}(r) + h_{\alpha{\rm Be}}(R),
\label{sepH}
\end{equation}
where the $\alpha$-$\alpha$ Hamiltonian, 
$h_{\alpha\alpha}(r)=T_r+V_{\alpha\alpha}(r)$, has a resonance at
$E_r^{\alpha\alpha}=92.08$ keV. The normalized 
wave function for the resonance is expressed as $\phi_r^{\alpha\alpha}(\vec r)$. 
In addition, a simple potential model is assumed for the $\alpha$-$^8$Be
relative motion, so that $h_{\alpha{\rm Be}}(R)=T_R+V_{\alpha{\rm Be}}(R)$.
The potential $V_{\alpha{\rm Be}}(R)$ is chosen so as to give a resonance at $E_r^{\alpha{\rm Be}}=287.7$ keV 
with the normalized wave function of $\phi_r^{\alpha{\rm Be}}(\vec R)$. 
The Hoyle state is then described by the wave function product, 
$\phi_r^{\alpha\alpha}(\vec r)\phi_r^{\alpha{\rm Be}}(\vec R)$, 
at the summed resonance energy of 
$E_r^{\alpha\alpha}+E_r^{\alpha{\rm Be}}=379.8$ keV.

The following approximation is then introduced.
The problem of potential scattering in either the $\alpha$-$\alpha$ 
or $\alpha$-$^8$Be system is considered. The resonance energy is denoted as $E_r$ 
and its normalized radial wave function as $u_r(r)$. 
Using $R$-matrix theory \cite{La58}, the radial wave function 
at an energy $E$ around a resonance with an asymptotic form of
$u_E(r) \rightarrow (2\mu/\pi\hbar^2k)^{1/2} \sin (kr+\delta)$
can be approximated by
\begin{equation}
u_{E}(r)=u_r(r) \sqrt{L(E,E_r,\Gamma_r(E))},
\label{uE_ap}
\end{equation}
where $L(E,E_r,\Gamma_r)$ is a Lorentzian function given by
\begin{equation}
L(E,E_r,\Gamma_r(E)) = \frac{1}{2\pi}
\frac{\Gamma_r(E)}{(E-E_r)^2 + \Gamma_r(E)^2/4}.
\end{equation}
Here the shift of the resonance energy is ignored. 
The energy-dependent width $\Gamma_r(E)$ is 
related to the width at the resonance energy $\Gamma_r$ by 
$\Gamma_r(E)=\Gamma_r P_l(E)/P_l(E_r)$, where $P_l(E)$ is 
the penetrability. 

Using Eq.~(\ref{uE_ap}), any function $f(H)$
of the three-body separable Hamiltonian $H$ in Eq.~(\ref{sepH}) can be approximated as
\begin{eqnarray}
&&f(H)=
\vert \phi_r^{\alpha\alpha}(\vec r) \rangle 
\langle \phi_r^{\alpha\alpha}(\vec r) \vert \cdot
\vert \phi_r^{\alpha{\rm Be}}(\vec R) \rangle 
\langle \phi_r^{\alpha{\rm Be}}(\vec R) \vert \cdot
\nonumber\\ && \times
\int dE_{\alpha\alpha} \int dE_{\alpha{\rm Be}}
L(E_{\alpha\alpha},E_r^{\alpha\alpha},
\Gamma_{r}^{\alpha\alpha}(E_{\alpha\alpha}))
\nonumber\\ && \times
L(E_{\alpha{\rm Be}},E_r^{\alpha{\rm Be}},
\Gamma_{r}^{\alpha{\rm Be}}(E_{\alpha{\rm Be}}))
f(E_{\alpha\alpha}+E_{\alpha{\rm Be}}),
\end{eqnarray}
where $\Gamma_{r}^{\alpha\alpha}$ and 
$\Gamma_{r}^{\alpha{\rm Be}}$ are the $\alpha$-decay widths of 
the $\alpha$-$\alpha$ and $\alpha$-$^8$Be resonances,
respectively. Substituting this into Eq. (\ref{rate-3a}) gives
\begin{eqnarray}
&&\langle \alpha\alpha\alpha \rangle
=
6 \cdot 3^{3/2} \left( \frac{2\pi\beta\hbar^2}{M_{\alpha}} \right)^3
\nonumber\\
&&\times
\int dE_{\alpha\alpha} \int dE_{\alpha{\rm Be}}
L(E_{\alpha\alpha},E_r^{\alpha\alpha},
\Gamma_{r}^{\alpha\alpha}(E_{\alpha\alpha}))
\nonumber\\
&&\times
L(E_{\alpha{\rm Be}},E_r^{\alpha{\rm Be}},
\Gamma_{r}^{\alpha{\rm Be}}(E_{\alpha{\rm Be}}))
e^{-\beta E_{\alpha\alpha}-\beta E_{\alpha{\rm Be}}} 
\nonumber\\
&&\times 
\Gamma_{\gamma}(^{12}{\rm C};0_2^+)
\left( \frac{E_{\alpha\alpha}+E_{\alpha{\rm Be}} -E(^{12}{\rm C};2^+)}
{E_r(^{12}{\rm C};0_2^+)-E(^{12}{\rm C};2^+)} \right)^5,
\label{Rmat_rate}
\end{eqnarray}
where $\Gamma_{\gamma}(^{12}{\rm C};0_2^+)$ is the 
radiative decay width of the Hoyle state and 
$E(^{12}{\rm C};2^+)$ is the excitation energy of the first 
$2^+$ state of $^{12}$C.
The rate expression given in Eq.~(\ref{Rmat_rate}) is almost equivalent 
to the empirical NACRE formula \cite{An99}. Thus, a formula quite close to the NACRE formula could be successfully derived, starting with a microscopic three-body Hamiltonian.


However, a question still remains regarding the validity of assuming that the three-body Hamiltonian is in fact separable. To resolve this, a numerical investigation was carried out to determine how much this assumption changes the calculated reaction rate.
It was found that using the separable Hamiltonian (\ref{sepH}) when solving equation (\ref{iteq}) changed the reaction rate by only a factor of two or less.
This indicates that if the Hamiltonian is constructed such that the $^8$Be ground state 
and $^{12}$C Hoyle state resonances are reasonably described, 
the separability approximation does not seriously affect the reaction rate.

The largest numerical difference between Eq.~(\ref{Rmat_rate}) and 
the NACRE expression is associated with the width $\Gamma_r^{\alpha{\rm Be}}$.
In the derivation used in the present paper, this quantity represents the decay 
width of the $\alpha$-$^8$Be resonance, while in the NACRE derivation it
is the particle decay width of the Hoyle state, which may 
include both $\alpha$-$\alpha$ and $\alpha$-$^8$Be decay. 
As previously stated, the reaction rate shown in 
Fig.~\ref{fig:rateNr} is smaller than the NACRE rate by 
a factor of up to six. It is considered that this is largely due to the difference in $\Gamma_r^{\alpha{\rm Be}}$.

Finally, the reason why different theoretical approaches yield such widely different reaction rates at low temperatures is considered.
In particular, although the CDCC approach \cite{Og09} 
employs almost the same model space as that adopted in the present paper, 
the result differs by 26 order of magnitude  at low 
temperature from both the NACRE rate and the rate obtained here. To clarify 
the origin of this discrepancy, the 
imaginary-time evolution of Eq.~(\ref{iteq}) was investigated using 
the coupled-channel method.

\begin{figure}[tb]
 \centering
 \includegraphics[scale=0.7]{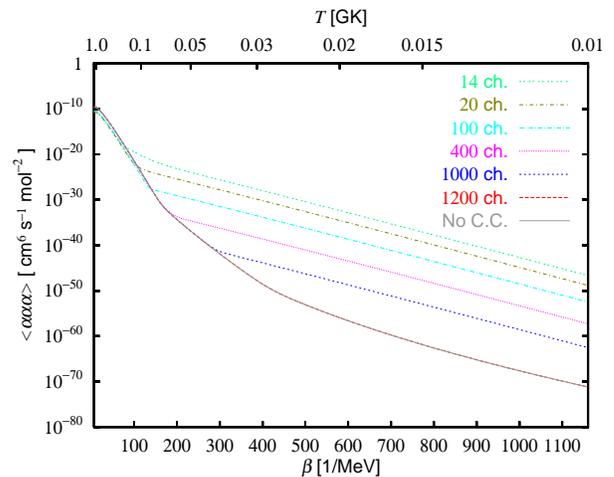}
 \caption{(Color online) 
Calculated triple-alpha reaction rate using coupled-channel expansion for different numbers of channels, $n_{\rm max}$.}
\label{fig:rateCCall}
\end{figure}

In the coupled-channel approach, the eigenvalue
problem for $\alpha$-$\alpha$ relative motion described by the
Hamiltonian $h_{\alpha\alpha}(r)$ is first solved. Discretizing the radial variable
$r$ in 0.5-fm steps up to 600 fm gives 1200 grid points for this
coordinate. Diagonalizing the radial Hamiltonian then gives
1200 eigenfunctions, $w_n(r)$ $(n=1-1200)$. Eigenstates 
associated with low eigenvalues are characterized by a large 
$\alpha$-$\alpha$ separation outside the Coulomb barrier,
except for the resonant state corresponding to the $^8$Be 
ground state, which appears as the 14th eigenstate. In the
coupled-channel approach, the wave function is expanded in the form 
$u(R,r,\beta)=\sum_n v_n(R,\beta) w_n(r)$ and 
the imaginary-time evolution of $v_n(R,\beta)$ is calculated in the form 
of a matrix differential equation. It was first numerically confirmed
that, employing all 1200 eigenstates in the expansion,
the calculated rate exactly matches the result shown 
in Fig.~\ref{fig:rateNr}, as is expected. However, if the number of basis functions in the expansion is truncated, the results depend on the degree of truncation. The dependence of the convergence behavior on the number of basis functions is shown 
in Fig. \ref{fig:rateCCall}. In each coupled-channel
calculation, the strength of the three-body potential
among the three $\alpha$-particles is adjusted so that the Hoyle state 
always appears at 379.8 keV.

Employing states below the $^8$Be ground state ($n_{\rm max}=14$), 
the reaction rate is much higher than that given by the fully converged 
calculation, and the difference is 24 orders of magnitude at $T=0.01$ GK. This can be easily understood because low-energy eigenstates 
of $h_{\alpha\alpha}(r)$ are characterized by a large separation
between two $\alpha$-particles outside the Coulomb barrier. 
The Coulomb barrier for the third $\alpha$-particle is then very 
small in these channels, yielding a high reaction rate. 
This artificial enhancement of the calculated reaction rate for a small channel number is reduced by off-diagonal coupling terms as the number of channels increases. However, as seen in Fig. \ref{fig:rateCCall}, convergence is extremely slow. Even with 400 channels,
the reaction rate is still 13 orders of magnitude larger than the
fully converged result. Since the $\alpha$-$\alpha$ energy for the
400th channel is around 46 MeV, which is far above the Coulomb barrier, 
the coupling effect relevant to the slow convergence is not 
related to physical distortion. It is considered that this indicates 
a difficulty in numerically expressing exponentially small functions 
in the tunneling process using the basis expansion method. 

In summary, imaginary-time theory was applied to determine
the triple-alpha reaction rate. Since the theory does not
require solving any scattering problems, it is quite
suitable for the triple-alpha process. Indeed, 
a converged reaction rate was obtained without any numerical problems.
The calculated rate agreed well with the conventional
NACRE rate, not only in terms of the magnitude, but also the critical temperatures
where the dominant reaction mechanism changes. No enhancement
of the rate was found at low temperature.
The reason for the good agreement was
analytically clarified using $R$-matrix theory. 
It was found that extremely slow convergence occurs if a coupled-channel expansion of the wave function is used, which helps to explain the very different reaction rates obtained using different theoretical approaches.


\begin{thebibliography}{99} 
\bibitem{Sa52} E.E. Salpeter, Astrophys. J. {\bf 115}, 326 (1952).
\bibitem{Ho54} F. Hoyle, Astrophys. J. Suppl. {\bf 1}, 121 (1954).
\bibitem{No85} K. Nomoto, F.-K. Thielemann, S. Miyaji, 
Astron. Astrophys. {\bf 149}, 239 (1985).
\bibitem{La86} K. Langanke, M. Wiescher, and F.-K. Thielemann, 
Z. Phys. A{\bf 324}, 147 (1986).
\bibitem{De87} P. Descouvemont and D. Baye, Phys. Rev. C{\bf 36}, 54 (1987).
\bibitem{An99} C. Angulo et.al, Nucl. Phys. {\bf A656}, 3 (1999).
\bibitem{Og09} K. Ogata, M. Kan, M. Kamimura, Prog. Theor. Phys.
{\bf 122}, 1055 (2009).
\bibitem{Ka86} M. Kamimura, M. Yahiro, Y. Iseri, Y. Sakuragi,
H. Kameyama, M. Kawai, Prog. Theor. Phys. Suppl. {\bf 89}, 1 (1986).
\bibitem{Do09} A. Dotter and B. Paxton, Astronomy and Astrophysics {\bf 507}, 1617 (2009).
\bibitem{Sa10} M. Saruwatari and M. Hashimoto, Prog. Theor. Phys. {\bf 124}, 925 (2010).
\bibitem{Su11} T. Suda, R. Hirschi, M.Y. Fujimoto,
Astrophys. J. {\bf 741}, 61 (2011).
\bibitem{Ga11} E. Garrido, R. de Diego, D.V. Fedorov, and A.S. Jensen,
Eur. Phys. J. A{\bf 47}, 102 (2011).
\bibitem{Is12} S. Ishikawa, Few-Body Syst. (2012).
\bibitem{Ng12} N.B. Nguyen, F.M. Nunes, I.J. Thompson, E.F. Brown,
Phys. Rev. Lett. {\bf 109}, 141101 (2012).
\bibitem{Is13} S. Ishikawa, Phys. Rev. C{\bf 87}, 055804 (2013).
\bibitem{Ng13} N.B. Nguyen, F.M. Nunes, and I.J. Thompson, Phys. Rev. C{\bf 87}, 054615 (2013).
\bibitem{Su13} Y. Suzuki, P. Descouvemont, arXiv:1308.4021.
\bibitem{Ya12} K. Yabana, Y. Funaki, Phys. Rev. C{\bf 85}, 055803 (2012).
\bibitem{Se92} T. Seideman and W.H. Miller, J. Chem. Phys. {\bf 96}, 4412 (1992);
{\bf 97}, 2499 (1992).
\bibitem{La58} A.M. Lane, R.G. Thomas, Rev. Mod. Phys. 30, 257 (1958).
\bibitem{de11} R. de Diego, E. Garrido, D.V. Fedorov, A.S. Jensen,
Phys. Lett. B{\bf 695}, 324 (2011).
\bibitem{de10}  R. de Diego, E. Garrido, D.V. Fedorov, and A.S. Jensen,
Europhys. Lett. {\bf 90}, 52001 (2010).
\bibitem{Al66} S. Ali and A.R. Bodmer, Nucl. Phys. {\bf 80}, 99 (1966).
\bibitem{OCM} T. Yamada and P. Schuck, Eur. Phys. J. A {\bf 26}, 185 (2005).
\bibitem{No82} K. Nomoto, Astrophys. J. {\bf 253}, 798 (1982).
\end{thebibliography}
\end{document}